\documentclass[onecolumn]{jpsj3}

\usepackage{xcolor}

\title{Benchmark test of Black-box optimization using D-Wave quantum annealer}

\author{Ami~S.~Koshikawa$^{1,2}$\thanks{amisk@dc.tohoku.ac.jp}, Masayuki~Ohzeki$^{1,\, 2,\, 3}$\thanks{mohzeki@tohoku.ac.jp}, Tadashi~Kadowaki$^{4}$, and Kazuyuki~Tanaka$^{1}$}
\inst{$^{1}$Graduate School of Information Sciences, Tohoku University, Sendai 980-8579, Japan\\
      $^{2}$Sigma-i~Co.,~Ltd., Tokyo 108-0075, Japan\\
      $^{3}$Institute of Innovative Research, Tokyo Institute of Technology, Yokohama 226-8503, Japan\\
      $^{4}$AI R\&I Division, DENSO CORPORATION, Tokyo 108-0075, Japan}

\abst{%
In solving optimization problems, objective functions generally need to be minimized or maximized.
However, objective functions cannot always be formulated explicitly in a mathematical form for complicated problem settings.
Although several regression techniques infer the approximate forms of objective functions, they are at times expensive to evaluate.
Optimal points of ``black-box'' objective functions are computed in such scenarios, while effectively using a small number of clues.  
Recently, an efficient method by use of inference by sparse prior for a black-box objective function with binary variables has been proposed. 
In this method, a surrogate model was proposed in the form of a quadratic unconstrained binary optimization (QUBO) problem, and was iteratively solved to obtain the optimal solution of the black-box objective function.
In the present study, we employ the D-Wave 2000Q quantum annealer, which can solve QUBO by driving the binary variables by quantum fluctuations.
The D-Wave 2000Q quantum annealer does not necessarily output the ground state at the end of the protocol due to freezing effect during the process.
We investigate effects from the output of the D-Wave quantum annealer in performing black-box optimization.
We demonstrate a benchmark test by employing the sparse Sherrington-Kirkpatrick (SK) model as the black-box objective function, by introducing a parameter controlling the sparseness of the interaction coefficients. 
Comparing the results of the D-Wave quantum annealer to those of the simulated annealing (SA) and semidefinite programming (SDP), our results by the D-Wave quantum annealer and SA exhibit superiority in black-box optimization with SDP\@.
On the other hand, we did not find any advantage of the D-Wave quantum annealer over the simulated annealing.
As far as in our case, any effects by quantum fluctuation are not found.
}
\begin{document}
\maketitle

\section{Introduction}
  Black-box optimization is a method to optimize complex and expensive intractable functions, and functions without derivatives or explicit forms. 
  Such functions appear in many problems in fields such as material informatics\cite{oganov2006}, machine learning\cite{snoek2012}, and robotics\cite{floreano1998}.
  A systematic way to perform black-box optimization is Bayesian optimization\cite{jones1998}.
  In this method, data points are randomly chosen to generate a training dataset for inferring the black-box objective function.
  A regression model is then constructed to predict a relation between the input variables and the black-box objective function in the training dataset. 
  Once the regression model is trained, an acquisition function is set up on its basis, which selects the next data point in a solution space from the trained model. 
  The optimal solution of the acquisition function is used to evaluate the black-box objective function, and to obtain a new data point of it.
  When this value is evaluated, the regression model is retrained with new data. 
  These steps are performed iteratively to pursue desired solutions, namely the optimal point of the black-box objective function.
  
  Bayesian optimization is applied mostly to black-box objective functions with continuous variables, because the optimization of the acquisition function is relatively straightforward.
  It may be applied to black-box objective functions with discrete variables as well.
  A significant bottleneck appears in problems with discrete variables, where the resultant acquisition functions also contain discrete variables.
  It is generally harmful to solve acquisition functions with discrete variables.
  Optimization problems with discrete variables often belong to the NP-hard class.
  It takes extremely long time to solve them using any algorithms.
  In a previous study, Bayesian optimization of combinatorial structures (BOCS)\cite{baptista2018} was proposed as a promising algorithm to evaluate the global minimum of black-box functions.
  In particular, a sparse prior was employed to efficiently perform regression in the Bayesian inference.
  The acquisition function was assumed as a quadratic unconstrained binary optimization (QUBO).
  Notably, relaxation to semidefinite programming (SDP) was used in the optimization phase, which can attain approximate solutions in a reasonable amount of time.

  Recently, D-Wave Systems Inc.\ developed a device\cite{johnson2011} that physically implements quantum annealing (QA)\cite{kadowaki1998}. It is a meta-heuristics to obtain the ground state of Ising spin glasses belonging to QUBO problems, and this device is now available commercially. 
  Because various combinatorial optimization problems can be formulated as Ising models\cite{lucas2014}, this D-Wave device has been used in the real world to solve a multitude of practical problems\cite{ohzeki2019, neukart2017, amin2018}. 
  The device uses niobium rings as quantum bits (qubits) with programmable local fields and mutual inductance of two qubits, so that the device can solve QUBO problems. 
  Solving a QUBO problem is equivalent to finding a ground state of an Ising spin glass, because binary variables can be rewritten as spin variables.
  The first stage of QA is initialized in the trivial ground state of the driver Hamiltonian.
  The quantum effect involved in the driver Hamiltonian is gradually turned off, and ends so that only the classical Hamiltonian with a nontrivial ground state remains.
  One of the standard choice of the driver Hamiltonian consists only of the $x$ element of the Pauli matrices called the transverse field.
  When the transverse field changes sufficiently slowly, the quantum adiabatic theorem ensures that we can find the nontrivial ground state at the end of QA\cite{Suzuki2005,Morita2008,Ohzeki2011c}.
  Numerous reports\cite{Santoro2002,Santoro2004} have shown that QA outperforms simulated annealing (SA)\cite{kirkpatrick1983}, which utilizes the thermal fluctuation and solves the combinatorial optimization problems.
  In context of machine learning, in which they solve various optimization problem in training, QA leads to a different kind of value in the output solution known as the generalization performance as in several literatures\cite{Ohzeki2018,Baldassi2018,arai2021}.

  A previous study on black-box optimization using the D-Wave device has used the factorization machine\cite{kitai2020}, which is used for recommendation systems and can be formulated in QUBO\@.
  They focused on metamaterial design, and evaluated the figure-of-merit in their metamaterial simulation.
  In the present study, we test the D-Wave quantum annealer in the black-box optimization by use of BOCS\@.
  In particular, the D-Wave quantum annealer does not necessarily output the ground state at the end of the procedure, partly because the connectivity realized in D-Wave 2000Q is a sparse structure called chimera graph. 
  To embed a desired graph expressing the structure of the problem on the chimera graph, redundant qubits with chain structures are used to enhance the connectivity. 
  This is because a single qubit possesses just six connections on average.
  We use a heuristic tool called minorminer\cite{cai2014} to embed the complete graph into the chimera graph. 
  Since qubits in the same chain must have the same up or down direction of their magnetic moments, interactions between the qubits are inferred as ferromagnetic interactions. 
  However, qubits in the same chain often do not have aligned magnetic moments, and this makes the solution undetermined. 
  We resolve these broken chains by a majority vote of the directions.
  This is one of the reasons why the performance of QA in D-Wave 2000Q is unreliable.
  To achieve better performance of QA in D-Wave 2000Q, various techniques were proposed previously\cite{Okada2019,Okada2019Potts,Okada2019Potts2}.
  In addition, several techniques avoid many interactions between variables are proposed\cite{Ohzeki2020,NTTdata2020}.
  The performance of the D-wave quantum annealer is affected by the freezing effect, which appears because of a lack of sufficient quantum fluctuations for driving binary variables at the last stage of QA\cite{Amin2015}.
  In addition, the thermal effect affects the dynamics of the spin variables as well as quantum fluctuation nontrivially\cite{Bando2020}.
  Thus, the output is generally deviated from the ground state, especially for the hard optimization problems.
  Therefore, several protocols employ a non-adiabatic counterpart beyond the standard protocol of QA\cite{Ohzeki2010a,Ohzeki2011,Ohzeki2011proc,Somma2012}, with the thermal effect\cite{Kadowaki2019}.
  
  In BOCS, the fully-connected Ising model is set as the acquisition function.
  In general, the model includes a hard optimization problem.
  Thus, the resulting solution from the D-Wave quantum annealer is not necessarily the ground state.
  The deviation from the ground state is expected to affect the performance of BOCS\@.
  We investigate the effect from the quantum device, while comparing the performance of the BOCS when optimizing the acquisition function by SA\@.
  It is worth noting that SA does not always yield the ground state of the acquisition function depending on schedule decreasing a control parameter, temperature.
  Although, in the protocol of QA, we also tune the transverse field to control the quantum fluctuation. 
  Therefore, the comparison roughly demonstrates the difference between thermal and quantum fluctuations.
  In the present study, we mainly focus on the performance of BOCS depending on the solver in the optimization phase of the acquisition function.
  We compare the results of BOCS by SDP, which leads to an approximate minimizer of the acquisition function and previously proposed in the original paper on BOCS\cite{baptista2018} as a solver in the optimization phase, to those by SA and QA\@.
  Also in the literature on BOCS, the performance employing SA in the optimization phase was investigated.
  It was not necessarily hard to solve the black-box objective functions, which were random spin systems including Sherrington-Kirkpatrick (SK) model with decaying interactions in distance measured in indices of spin variables, which is essentially one-dimensional Ising spin glass with long-range interactions.
  In the original paper, the superiority of BOCS by SDP compared to one by SA was reported.
  However, it is not enough to discuss the performance of BOCS by solving the problems appeared in the previous study.
  In the present paper, we change the problem setting into harder one, sparse SK model as the black-box objective function.
  In this sense, the setting is completely different from the original study.
  In addition, we utilize the D-Wave quantum annealer to perform the optimization in BOCS not only by SA in a classical computer because this is also a candidate of the solvers employed in BOCS.\@
  
  In the procedure of BOCS, we have no information on the interaction strengths of the black-box objective function. In BOCS, we iteratively find lower-energy state of the acquisition function as a candidate of the optimal solution of the black-box objective function, while the coefficients in the acquisition function are changed.
  
  In order to investigate the performance of the BOCS from a different perspective, we consider the case when the form of the black-box objective function is known.
  Then we may perform regression to infer only the coefficients to reveal the objective function.
  The inferred coefficients lead to a good approximation of the black-box objective function.
  Then, we optimize the resulting approximate function and attain the good estimator of the minimizer of the objective function.
  In a previous study, the regression method was proposed only from pairs of spin configuration and the corresponding energy value\cite{takahashi2018,Ohzeki2018NOLTA}. 
  The method works particularly well for Ising spin glass with sparse interactions.
  An analytical study using a sophisticated replica method and a numerical verification revealed a relationship between the number of the zero-value coefficients and the number of data needed to reconstruct all the coefficients. 
  If the interaction matrix is sparse, fewer data are needed than the number of coefficients. 
  In BOCS, similar to this regression method, we utilize the sparse prior distribution to infer the coefficients of the surrogate model, but the form of the objective function is unknown.
  In the present study, we set the SK model as the black-box objective function.
  We assume that the surrogate model takes the same form as the black-box objective function.
  In other words, the present case is that we know the form of the black-box objective function.
  The comparison of the BOCS to the regression and optimization clarifies the performance to infer the sparse interactions of the black-box objective function.
  We measure the efficiency of the BOCS for the sparse SK model in terms of the necessary number of data, which corresponds to the number of iteration in BOCS, to find the optimal solution of the black-box objective function.
  In the previous study, although the BOCS was proposed as the black-box optimization technique for sparse interactions in the objective function, the performance of the BOCS depending on the sparseness remained unclear.
  To solve this problem, we change the sparseness of the SK models, and focus on the necessary number of data before finding the ground state by investigating the success rates in finding a minimum.
 
  The remainder of this paper is organized as follows: we explain the BOCS method in Sect. 2, we discuss the numerical experiments in Sect. 3, we compare the performances between the BOCS and regression with a sparse prior in Sect. 4, and summarize in Sect 5.
  
  \section{Method}
  We assume that the input $\vec{x}$ is a vector of binary variables, its $i$th element is denoted by $x_i$, and $N$ is the dimension of $\vec{x}$. 
  Each $\vec{x}$ provides an observation $y$ containing a finite error $\sigma$.
  Our goal is to find $\vec{x}$ that minimizes a black-box function. 
  Since we cannot determine an explicit form of the black-box function, we employ a surrogate model and train it using the values of the black-box objective function.
  Any objective function on the $N$-dimensional binary variables can be expressed by using up to the $N$-th polynomial of $\vec{x}$, though this polynomial requires $\mathcal{O}(2^N)$ data to fix all the parameters. 
  This huge amount of data cannot be collected in practical situations.
  We may thus cut the polynomials in finite orders.
  When the surrogate model contains higher-order terms than quadratic ones, we do not optimize it efficiently in general.
  In addition, the approximate surrogate model up to the second-order terms can be solved by SDP or the D-Wave quantum annealer efficiently in the optimization phase.
  We thus set a quadratic model as the surrogate model:
  \begin{equation}\label{eq:acquisition}
      \tilde{f}(\vec{x}) = \alpha_0 + \vec{x}^\top Q_\alpha \vec{x}.
  \end{equation}
 where $\alpha_0$ is a real value, and $Q_\alpha$ is an upper triangular matrix.

  In this algorithm, we compute a posterior distribution for the model parameters by following the framework of Bayesian inference.
  By sampling from the posterior distribution, we construct an acquisition function that indicates the next data point to choose from a solution space.
  We find the minimum of this acquisition function by using some optimization solvers. 
  Then, we obtain the new data from the black-box function with the $\vec{x}$ value. 
  This minimizes the acquisition function. 
  We retrain the model with data, including the new value from the black-box objective function. 
  This algorithm iteratively searches for the global minimum by updating the data points.
  Figure 1 shows a schematic drawing of the BOCS algorithm.
  \begin{figure*}[t]
      \centering
      \includegraphics[scale=0.45, pagebox=cropbox]{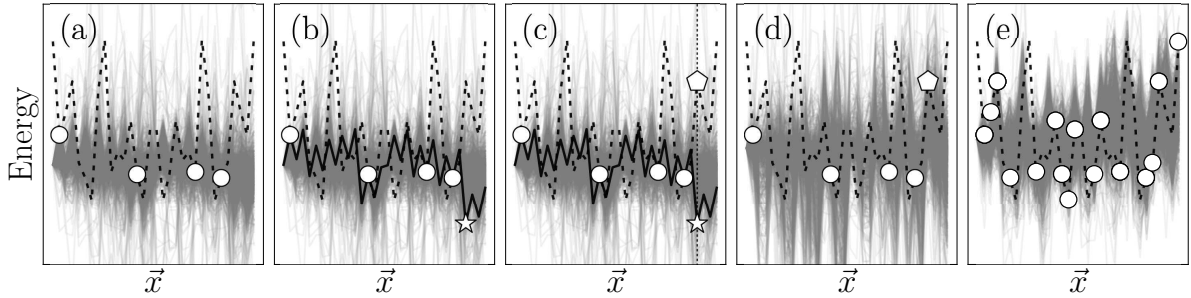}
      \caption{Schematic drawing of BOCS algorithm. 
      (a) Evaluation of a black-box function (dashed line) at four data points (open circles), and training a surrogate model with the four data points. The grey lines show surrogate models with regression parameters sampled 1000 times.
      (b) Construction of an acquisition function (solid line) by sampling a regression parameter from the posterior distribution. 
      An open star represents the optimal solution of the acquisition function.
      (c) Evaluation of the black-box function at the new data point (open pentagon).
      (d) Retraining the surrogate model by using the five data points (open circles and open pentagon).
      (e) Trained surrogate model with 16 data points (open circles).
      }\label{fig:regression} 
  \end{figure*}
  
\subsection{Constructing the Acquisition Function}
  We assume a few data points are attained, because the evaluation of objective functions is expensive. 
  We then consider the sparse Bayesian linear regression to take uncertainties of the regression parameters $\vec{\alpha} = [\alpha_0,\ Q_{\alpha 11},\ Q_{\alpha 22},\ \dots,\ Q_{\alpha 12},\ Q_{\alpha 13},\ \dots]$ 
  and observation noise $\sigma$. 
  From observations of several data points $\{\vec{x}^{(i)}, y^{i} \}_{i=1, 2,\dots}$, we compute a posterior distribution over $\vec{\alpha}$ as: 
\begin{align}
    \mathcal{P}(\vec{\alpha} |\vec{y}, X) \propto \mathcal{P}(\vec{\alpha})\mathcal{P}(\vec{y} | X, \vec{\alpha})\quad X \in \{ 0, 1 \}^{p\times D}
\end{align}
where we construct a matrix $X$ from the vector $\vec{x}^{(i)}$ as $\vec{X}^{(i)} = [1, x_1^{(i)}, x_2^{(i)},\dots, x_1^{(i)}x_2^{(i)}, x_1^{(i)}x_3^{(i)}, \dots]$.
In addition, $p = 1 + N + N(N - 1) / 2$ and $D$ represents the number of data points. 
Then, we set a likelihood function and a prior distribution
  over the parameter $\vec{\alpha}$. 
  The likelihood function is given by a Gaussian distribution with a variance $\sigma^2$ as:
\begin{align}
    \mathcal{P}(\vec{y}|X, \vec{\alpha}) = \mathcal{N}(\vec{\alpha} X, \sigma^2 I)
\end{align}
  Since the number of the elements of $\vec{\alpha}$ is $\mathcal{O}(N^2)$, the amount of data also needs $\mathcal{O}(N^2)$ to estimate the regression parameters. 
  Otherwise, we will obtain high-variance estimators. 
  To avoid getting uncertain parameters even when the data are scarce or the input dimension is large, we set a prior distribution.
  We here use a horseshoe distribution as the prior distribution to omit the hyperparameters to perform the BOCS.\@
  This prior distribution is capable to efficiently infer sparse parameters in the model even if the number of data is small\cite{Carvalho2010}: 
\begin{align}
    \alpha_k |\beta_k^2,\, \tau^2,\, \sigma^2 &\sim\mathcal{N}(0, \beta_k^2, \tau^2, \sigma^2)\quad k=1,\dots,p \nonumber \\
    \tau, \beta_k &\sim \mathcal{C}^+ (0, 1)\quad k = 1,\dots,p \nonumber \\
    \mathcal{P}(\sigma^2) &= \sigma^{-2},
\end{align}
  where $\mathcal{C}^+(0, 1)$ is the standard half-Cauchy distribution.
  This formulation, however, cannot realize efficient sampling. 
  Following Makalic and Schmidt\cite{makalic2016}, the half-Cauchy distribution can be expressed with the inverse-Gamma distribution to introduce auxiliary parameters. 
  The inverse-Gamma distribution is a conjugate prior for normal distribution. 
  The modified formulation is a closed-form from which we can efficiently sample parameters from this distribution with complexity $\mathcal{O}(p^3)$. 
  We use a faster algorithm\cite{bhattacharya2016} ($\mathcal{O}(D^2 p)$) that is exactly the same as that in the formulation with auxiliary parameters.

  As tested in the previous study, one may compute $\vec{\alpha}_\textit{MLE}$ by using a maximum likelihood estimation.
  Then BOCS with $\vec{\alpha}_\textit{MLE}$ showed purely exploitative behavior and failed to evaluate the optimal solution. 
  In the present study, we set coefficients $Q_\alpha$ by sampling from the posterior distribution $\mathcal{P}(\vec{\alpha} |X, \vec{y} )$, so that the BOCS algorithm shows exploring behavior in the solution space. 
  This is inspired by Thompson sampling in the context of bandit problems, which often shows better performance of the surrogate model attained by the maximum likelihood estimation.
  
  Notice that the horseshoe prior efficiently estimates the sparse parameters in the surrogate model.
  Thus the above formulation of BOCS exhibits better performance for the case in which the black-box objective function inherently the sparse interactions when we write its explicit form in a quadratic form.
  Similarly, as a prior distribution, we may use the Laplace distribution, which is typically chosen for estimating the sparse parameters.
  Due to existence of the hyperparameter in the Laplace distribution, the fair performance comparison between the Laplace distribution and horseshoe distribution is difficult in general. By tuning the hyperparameter, the result on the Laplace distribution can change and may become closer to the result from horseshoe distribution. 
  The superiority of the sparse prior can be expected from sharpness of the shape of the prior distribution.
  From this point, the horseshoe prior has a nice property to infer the sparse parameters compared to the Laplace distribution.

\subsection{Optimization Solver}
  Once $Q_\alpha$ is fixed, we optimize the acquisition function (\ref{eq:acquisition}) to select the new evaluation point. 
  The main contributor of this study is D-Wave 2000Q quantum annealer that solves discrete quadratic problems as well as SA and SDP\@. 

  Note that it can solve only $64$-variable problems on a complete graph due to sparsity of the hardware graph on the quantum processing unit in the D-Wave quantum annealer.
  The current quantum annealer (D-Wave advantage) has $5000+$ qubits and it implements the $180$ variables on a complete graph.
  
  Although the solvable size of the D-Wave quantum annealer is limited, the natural computation performed in the D-Wave quantum annealer following the protocol of QA outputs a near-optimal solution in relatively short time about 20 microseconds.
  In this sense, we may expect that the D-Wave quantum annealer can be a fast solver of QUBO\@.
  In general, it takes a relatively long time to solve the exact solution of QUBO as in explanation of SDP\@. 

  We explain the three solvers used in this study: SA, SDP, and QA by D-Wave 2000Q\@.
  
\paragraph{Simulated annealing}
  SA is a meta-heuristic utilizing thermal fluctuations in computation. 
  A spin configuration corresponding to binary variables starts 
  from a random state in a solution space, and one spin in the configuration is flipped following the Metropolis-Hastings algorithm\cite{Hasting1970}. 
  The energy difference between the initial state and the one-spin flipped state is denoted by $\Delta E$. 
  If $\Delta E < 0$, the state is updated to the flipped one, otherwise updated with a probability $\mathrm{e}^{-\Delta E/T}$. 
  The parameter $T$ is diminished in every iteration. 
  When $T$ is large, SA updates the spin configuration regardless of 
  $\Delta E$, and the system moves in a wide range in the solution space. 
  When $T$ becomes smaller, it magnifies the energy landscape and fall into the local minimum, because the state will not update without small $\Delta E > 0$ values.
  SA can thus be trapped into a local minimum in general.
  When the speed to control the temperature is very slow, SA can attain the ground state of the system, namely the optimal solution.
  Practically, a relatively quick sweep of the temperature can lead to the optimal solution in most cases.
  
\paragraph{Semidefinite programming}
  We briefly describe the application of SDP in solving discrete optimization problems. 
  We consider the following quadratic constrained problem in general as follows: 
  \begin{align}\label{eq:vector}
      \mathrm{minimize}_{\vec{y}} \quad &\sum_{i, j} C_{ij} y_i y_j + d \nonumber \\
      \mathrm{subject\ to}\quad &\sum_{i, j}D_{ijk} y_i y_j = b_k, \quad k=1, 2,\dots, K \\
      &\vec{y} \in \mathbf{R}^n \nonumber
  \end{align}
  Here $y_i y_j$ can be regarded as the $(i,j)$-element of the Gram matrix, whose eigenvalues are non-negative.
  This problem can be thus transformed to a SDP as 
  \begin{align}
      \mathrm{minimize}_X\quad &{\rm Tr} \left( C X \right) + d \nonumber \\
      \mathrm{subject\ to}\quad &{\rm Tr}\left( D_k X \right) = b_k, \quad k =1, 2,\dots, K \\
      &X \succeq 0 \nonumber
  \end{align}
  Here $X \succeq 0$ denotes that $X$ is a semidefinite matrix, which has non-negative eigenvalues. 
  The resultant minimization problem is a convex optimization problem.
  Thus we readily attain the optimal solution.
  The minimum value of this convex optimization problem is the same as that of the original one.
  We use the property of SDP to attain an approximate solution of our acquisition function. 
  The optimization of the acquisition function is as follows:
  \begin{align}
      \arg\min_{\vec{x}} \tilde{f}(\vec{x}) &= \arg\min_{\vec{x}}\left(\sum_i Q_{\alpha ii}x_i + \sum_{i<j} Q_{\alpha ij}x_i x_j\right) \nonumber \\
      &= \arg\min_{\vec{x}} \left(\vec{a}^\top + \vec{x}^\top A\vec{x}\right).
  \end{align}
  We then replace each binary variable $x_i$ with $\sigma_i = 2x_i - 1$, and the minimization problem can then be written as:
  \begin{align}\label{eq:combi}
      &\arg\min_{\vec{\sigma^\prime}} \vec{\sigma}^{\prime\top} C \vec{\sigma}^\prime,\quad \vec{\sigma}^\prime = [\vec{\sigma}^\top, \sigma_0]^\top \in \{-1,\, 1\}^{N+1} \nonumber \\
      &C = \begin{bmatrix} 
          \tilde{A} & \vec{c} \\
          \vec{c}^\top & 0
      \end{bmatrix} \nonumber \\
      &\tilde{A} = A / 4,\quad \vec{c} = \vec{a} / 4 + A^\top \vec{1} / 4.
  \end{align}
  We relax the binary variable $\sigma_i$ to a vector $y_i$ on the $N+1$-dimensional unit sphere.
  We can then rewrite eq. (\ref{eq:combi}) as eq. (\ref{eq:vector}) with $d = 0$, $D_{ijk} = \delta_{ij}\delta_{ik}$, $\vec{b} = \vec{1}$ and $k = N+1$.
  We instead solve the resultant SDP as a relaxation problem to generate the approximate solution.
  We binarize the attained approximate solution using an adequate binarization technique according to the previous study\cite{baptista2018}.
  In general, it takes an exhaustive time to solve the original optimization problem with binary variables.
  However, in BOCS, we must iteratively optimize the acquisition function.
  We should thus employ approximate techniques for performing the BOCS in a reasonable amount of time.
  
\paragraph{D-Wave 2000Q quantum annealer}
  D-Wave 2000Q is a commercial quantum annealer from D-Wave Systems Inc., which physically implements the Ising model with the transverse field. 
  By mapping combinatorial optimization problems into the two-body Ising model, it is possible to find near-optimal solutions based on QA in a few microseconds. 
  In particular, at the end of QA in the D-Wave quantum annealer, the weak quantum fluctuations by the transverse field cannot drive the spins. 
  This is known as the freezing phenomena\cite{Amin2015}.
  Thus, the spin configuration often deviates from the ground state.
  
  In addition, the connectivity realized in D-Wave 2000Q is a sparse structure called chimera graph. 
  We use a heuristic tool called minorminer\cite{cai2014}, to embed the complete graph 
  into the chimera graph. 
  Redundant qubits are formed into chain structures to realize a larger graph connectivity. While the magnetic moments of redundant qubits in a chain structure should take the same direction, they often take different directions. To obtain a solution with these redundant qubits, the direction is adopted by a majority vote.
  This is also a reason why the performance of QA in the D-Wave 2000Q gets worse on dense graph problems.
  Another postprocess to find the better solution by fixing the broken chain is minimizing energy.
  This technique is often find the lower energy state than the results after majority vote.
  In the present study, we choose the majority vote to perform the benchmark without any further improvements from the default setting of using the D-Wave 2000Q.
  
  Although there are a few reasons spoiling the output from the D-Wave 2000Q differing from the ground state, it can often yield the ground state in various types of optimization problems described in QUBO with a small number of variables.
  
\section{Numerical Experiment}
  We perform numerical experiments employing the SK model as a black-box objective function. 
  The SK model belongs to the NP-hard problem depending on the parameters 
  described by spins $\vec{\sigma} \in \{-1, 1\}^N$ and interactions $J_{ij}$:
\begin{align}
    \mathcal{H} = -\frac{1}{N} \sum_{i<j} J_{ij} \sigma_i\sigma_j.
\end{align}
The interaction coefficients are randomly selected as:
\begin{align}
    J_{ij} \sim (1-\rho)\mathcal{N}(0, +0) + \rho\mathcal{N}(0, 1).
\end{align}
  The standard definition of the SK model is that the coefficient $1/N$ should be $1/\sqrt{N}$.
  However, we introduce a parameter to control sparseness $\rho \in (0, 1]$,
and set the coefficient as $1/N$.
  If the parameter $\rho$ is close to zero, 
  the number of zero-elements in the $J_{ij}$ matrix increases.
  BOCS is expected to work well, because it implements the sparse prior.
  The performance of BOCS should be discussed in two ways.
  The first is inference.
  The value of $\rho$ affects the performance of inference in BOCS\@.
  This is independent of the optimization solvers.
  The other way is in optimization.
  The sparse connectivity of the Ising spin glass set in the black-box objective function affects the performance of the optimization solvers.
  In this study, we use SDP to quickly generate the approximate solution of the acquisition function.
  SA and QA on the D-Wave quantum annealer do not always reach the ground state but directly solve the acquisition function.
  Notice that we perform iterations in SA in fixed steps during the optimization phase.
  In addition, QA on the D-Wave quantum annealer performs only in a fixed amount of time.
  
  In this study, we treat SK models with $N=20$ spins, and $10$ initial datasets $\{\vec{x}^{(i)}, y^{(i)}\}_{i=1,\dots,10}$.
  The parameter $\rho$ varies from $0.1$ to $1.0$ in the increment of $0.1$, and we investigate its dependence on performance. 
  For every $\rho$ value, we generate $50$ instances and compute each problem for $10$ runs.

\begin{figure*}[t]
     \centering
     \includegraphics[scale=0.7]{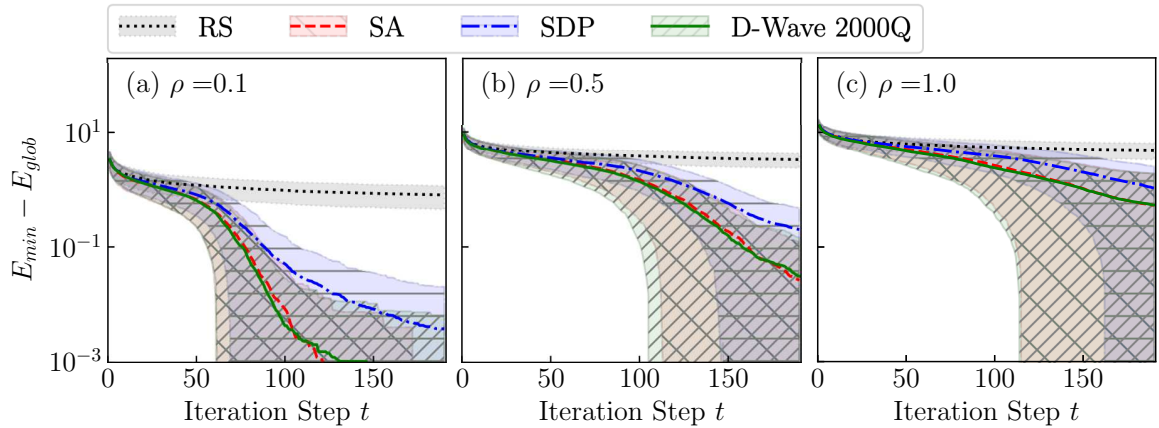}
     \caption{(Color online). Subtraction of the global minimum ($E_\textit{glob}$) from the minimum value in the dataset ($E_\textit{min}$) at $\rho=0.1$ (a), 
     $\rho=0.5$ (b), and $\rho=1.0$ (c). Each curve represents the average of all trials,
     and each hatch stands for corresponding standard deviations. 
     The solid curve, the dashed curve, 
     and the dash-dotted curve respectively indicate the result of BOCS with D-Wave 2000Q, SA, and SDP. }%
     \label{fig:E-vs-step}
\end{figure*}

  Figure~\ref{fig:E-vs-step} shows the residual energy after $t$ BOCS iteration steps at $\rho=0.1$ (a), $\rho=0.5$ (b), and $\rho=1.0$ (c).
  The residual energy is calculated by subtracting the global minimum through brute-force computation ($E_\textit{glob}$) from the minimum value in the obtained data ($E_\textit{min}$). 
  The curves indicate the mean values of all trials, and the shaded areas indicate the standard deviation. 
  Each curve and hatch type refer to solvers used in the optimization phase: the solid, the dashed, and the dashed-dotted curves show results from D-Wave 2000Q, SA, and SDP, respectively. 
  In addition, the dotted curve shows the result of random search (RS), which randomly chooses a data point at each iteration step.
  The performance of BOCS with any optimization solver is superior to that of RS\@. 
  In our SK models, there is no significant difference between using SA and D-Wave 2000Q, while BOCS with SDP does not efficiently decrease the resulting energy.
  This is contrast from the previous result in the original paper of the BOCS\cite{baptista2018}.

  The D-Wave 2000Q generates a lower-energy state at the end of the protocol.
  The lower-energy state is governed by the Gibbs-Boltzmann distribution with a finite value of the transverse field.
  In a sense, the resultant spin configuration is affected by the quantum fluctuation.
  Furthermore, SA in a finite number of steps remains in the thermal fluctuation in its resultant solutions.
  Therefore, the comparison between the results by SA and D-Wave focuses on the difference between the thermal and quantum fluctuations.
  However, we did not find difference between thermal and quantum fluctuations in our experimental setup.

  We hereafter focus on the comparison between the results by SDP and SA\@.
  The difference between the two cases is mainly the deviation from the tentative ground state of the acquisition function.
   We simply assume that the BOCS by SDP falls into a local minimum in the acquisition function at each step, which explains why the BOCS by SDP gets worse performance.
   In the BOCS the balance between exploration and exploitation of the Bayesian inference is important.
   As we introduce the sampling technique in BOCS for increasing the exploratory property inspired by the Thompson sampling, the BOCS by SDP sometimes approach the optimal solution.
  However, the SA (and D-Wave) outperforms SDP in our problem setting, due to the better performance to attain the lower-energy state by SA (and D-Wave) of the acquisition function.
  In our problem setting, we employ the sparse SK model.
  Thus the acquisition function also takes the similar spin glass problem during the process of BOCS\@.
  This is one of the reasons why SDP shows worse performance compared to SA (and D-Wave).
  
  The difference between SDP and SA (and D-Wave) becomes small as the $\rho$ value increases. 
  This implies that the exploratory space gets narrower as $\rho$ takes higher values.
  In other words, the solution space is divided into the states around many deep local minima, and thus, the exploratory space cannot be sufficiently broadened even by using thermal or quantum fluctuations.
  The difficulty in solving the hard problems appear in the performance in BOCS\@. 
  
\begin{figure*}[t]
    \centering
    \includegraphics[scale=0.6]{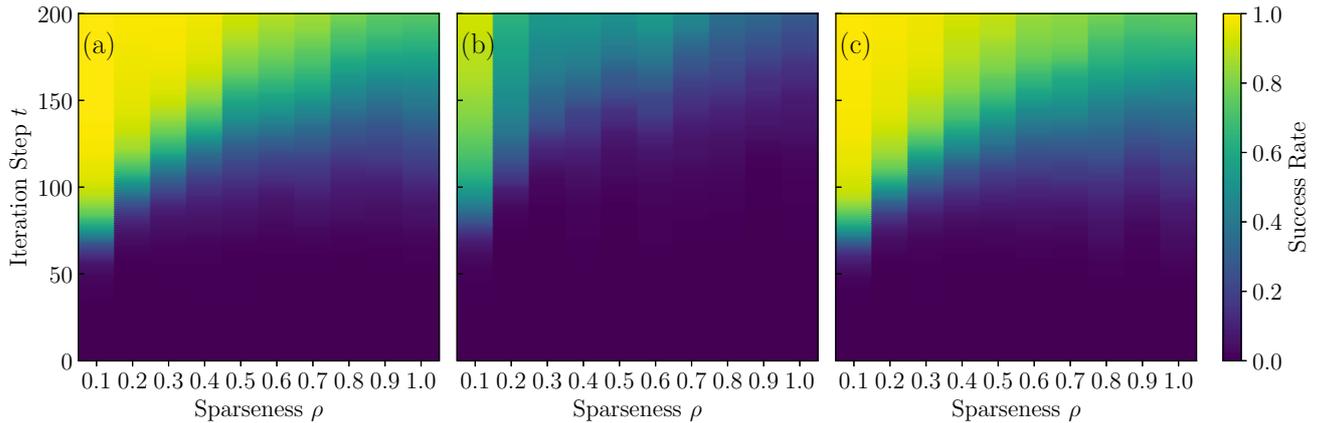}
    \caption{(Color online). Success rate of finding the global minimum as a function of $\rho$ and the iteration step when SA (a), SDP (b),
    or D-Wave 2000Q (c) is used.}\label{fig:step-vs-rho}
\end{figure*}

  In order to investigate the dependence of the performance of BOCS on $\rho$, we plot a success probability of finding the global minimum in Figure~\ref{fig:step-vs-rho}.  
  Regardless of the optimization solvers used, the number of iteration steps before finding the minimum value becomes large as $\rho$ increases.
  The number of iteration steps consists of the number of data points used in BOCS and the number of duplication.
  Moreover, we replace the number of iteration steps with the number of data points as shown in Figure~\ref{fig:data-vs-rho}.
  Then, we find that  the same dependence on $\rho$ as that in Figure~\ref{fig:step-vs-rho}.

\section{Comparison with Regression}
  If the form of the black-box objective function is known a priori while its coefficients in the quadratic form and some parameters are unknown, 
  one may infer only the coefficients  from the regression data attained. 
  This means that once the required number of data points is collected at random, one can reconstruct the coefficients,
  and the minimum solution can be obtained with an appropriate optimization solver. 
  To reconstruct  the coefficients, we solve the following equations:
\begin{align}
    \min_{\tilde{\boldsymbol{J}}} \| \tilde{\boldsymbol{J}} \|_1\quad \mathrm{subject\ to}\ \boldsymbol{E} - S\tilde{\boldsymbol{J}} = \boldsymbol{0},
\end{align}
  where $\tilde{\boldsymbol{J}}$ is the coefficient vector, its element is $J_{ij}$, $S$ is the spin-data matrix, the $i$th spin data vector is denoted by $S^{(i)} = [\sigma^{(i)}_1\sigma^{(i)}_2, \dots, \sigma^{(i)}_{N-1}\sigma^{(i)}_N],\ S \in \{0,\, 1\}^{D \times N(N-1)/2}$ and $E$ is the energy-data vector.
  Here $E^{(i)}$ equals to the energy of a spin configuration $S^{(i)}$.
  In the literature\cite{takahashi2018}, the replica method revealed the relationship between $\rho$ and the required number of data points that reconstructs $J_{ij}$. The analysis was validated by numerical experiments. 
  To compare the performance by BOCS and the previous results\cite{takahashi2018}, 
  we convert the iteration steps in Figure~\ref{fig:step-vs-rho} into the number of obtained data points. 
  BOCS performs a random search at the beginning, because its surrogate model has coefficients with large uncertainties, and the acquisition function varies drastically at each step. 
  When the data are collected partially, the acquisition function moderately changes.
  Therefore, new data may not always be obtained at each step, and previously obtained data may be selected instead.
  The number of iteration steps is not equivalent to the number of data points. 
\begin{figure*}[t]
    \centering
    \includegraphics[scale=0.6]{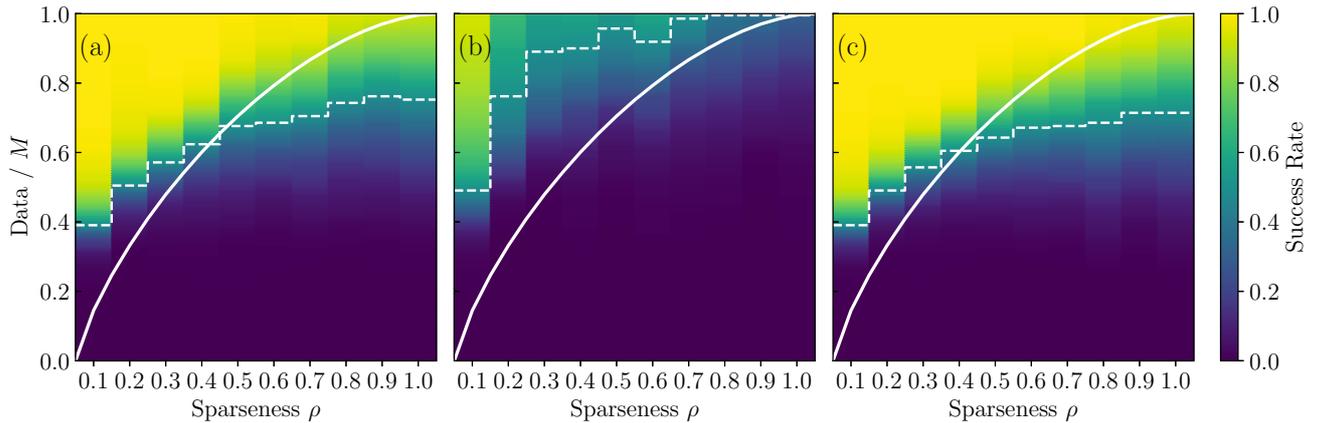}
    \caption{(Color online). Success rate of finding the global minimum as a function of $\rho$ and the number of data points divided by the number of coefficients
    when SA (a), SDP (b), or D-Wave 2000Q (c) is used.
    The solid curve shows the results of the replica analysis. The dashed line shows where the success rate is 50\% when BOCS is used.
    The lower curve has a better performance for a given $\rho$. The performances of SA and D-Wave 2000Q are insensitive to the value of $\rho$ compared to the results of the previous study.
    }\label{fig:data-vs-rho}

\end{figure*}
  Figure~\ref{fig:data-vs-rho} shows the success rate as a function of $\rho$ where the number of data points that are divided by the number of $J_{ij}$ parameters.
  The heat map indicates the success rate for obtaining the minimum value. 
  The dashed line indicates a border where the success rate is 50\% by using BOCS\@.
  The solid white curve is the previous results in the literature\cite{takahashi2018}. 
  The area above this line indicates that the number of obtained data points is large enough to reconstruct $J_{ij}$ by regression. The other area indicates the number of data points that cannot reconstruct the $J_{ij}$ and thus, this solid curve can be regarded as a phase transition line.
  Since the results given by the previous study are typical reconstruction limits, in the case where $N \to \infty$, the success and the failure area are clearly separated. 
  However, our numerical experiments of BOCS are for finite-number tests.
  Therefore, the borders of the success area are rather ambiguous. 
  If SA (and D-Wave) is used for the optimization solver in a complicated problem, BOCS is more likely to find the minimum value with a relatively small number of data points. 
   Under certain situations, the BOCS occasionally yields better performance than the typical reconstruction limit that was revealed in the previous study.
  We propose a few reasons that explain this observation.
  The first is that the BOCS only focuses on finding the ground state.
  The regression does not directly derive the ground state.
  The typical reconstruction limit is the performance of inferring the correct coefficients of the black-box objective function.
  Once we find the correct coefficients, we can get the exact ground state of the black-box objective function.
  One might find the ground state from approximate values of the coefficients when the number of data points is insufficient.
  However, below the typical reconstruction limit, a drastic change in the coefficients appears compared to the correct coefficients, because the system undergoes phase transition.
  Therefore, we cannot optimistically expect that the ground state of the black-box objective function is attained.
  Another reason is due to the difference between the sparse priors.
  In the previous study, the $L_1$ norm was used for inference of the sparse coefficients
  In addition, the hyperparameter for the Laplace distribution was, in some sense, optimized in their analysis.
  However, BOCS utilizes the horseshoe distribution, which may infer the sparse coefficients more efficiently.
  
  Notice that the results seem to be beyond the theoretical reconstruction limit as $\rho > D/M$.
  This would be a finite-size effect.
  We are not arguing that our results suggest any advantage of BOCS beyond the theoretical reconstruction limit in the region where $\rho$ takes a relatively large value.
  That being said, the possibility remains that BOCS, by making use of the horseshoe prior, reaches the theoretical reconstruction limit.
  We will be investigating this further in a future study.
 
\section{Summary and Future Directions}
  Black-box optimization aims at reducing the value of objective functions that are expensive to evaluate, and has broad applications in fields such as machine learning and robotics. 
  In the present study, we tested BOCS by setting the SK model as a black box objective function, and evaluated its minimum value. 
  In the optimization phase of BOCS, we proposed using the D-Wave 2000Q quantum annealer, which is expected to return near-optimal solutions in constant time, regardless of its problem size up to the limit of the capacity to embed the problem. 
  In particular, the D-Wave quantum annealer outputs the lower-energy state affected by a finite strength of the quantum fluctuation.
  Similar to the D-Wave quantum annealer, we use SA in a finite number of steps.
  The comparison between SA and the D-Wave quantum annealer clarifies the effects of the thermal and quantum fluctuations in BOCS\@.
  BOCS iteratively evaluates the tentative acquisition function.
  Thus, we expected that both thermal and quantum effects were present in the results of BOCS\@.
  Despite our hypothesis, we did not find difference between the thermal and quantum fluctuations in using BOCS\@.
  In addition, we also employed SDP in the optimization phase of BOCS\@. 
  The results by SA (and D-Wave) showed better performance than those of SDP in the SK model. 
  This is possibly due to the degree of deviation from the ground state of the tentative acquisition function.
  
  We also compare the number of required data points to find the minimum value of the black-box objective function by BOCS and regression with the $L_1$ norm. 
  Although BOCS needs more data points than regression to obtain the minimum value in most of the cases,
  there is a possibility of finding the minimum value with a smaller number of data points than in regression with the $L_1$ norm, when the black-box has a dense structure.
  Although the $L_1$ norm is used to perform regression to infer the sparse parameters, we employed the horseshoe prior in BOCS\@.
  This is possibly explained by the difference of the priors used, and will be investigated in a future study.
  We emphasize that our problem setting is slightly different from regression under the assumption of the form of the black-box objective function.
  We search for the minimum without knowledge on the details of the structure of the cost function.
  In addition, our results suggest that our approach can find only the minimum more efficiently, compared to the case of the regression with the $L_1$ norm finding the parameters to express the objective function.
  In this study, we set the SK model as the black-box objective function, which is of the same form as the acquisition function.
  In other words, the black-box objective function can be expressed by the acquisition function in principle.
  The performance of BOCS has not been sufficiently investigated when the black-box objective function is not of the same form as the acquisition function.
  We will also investigate this further.
  In addition, notice that, although we here choose the SK model as the benchmark test, the application of BOCS is not restricted on the objective function with two-body interactions.
  Beyond two-body interactions, the BOCS is available in principle.
  The performance of BOCS diminished by mismatch between the objective function and surrogate model will be the next scope along the same line as the present study.

  One of the reasons to employ the D-Wave quantum annealer as the optimization solver is utilizing quantum fluctuations.
  In order to investigate the quantum fluctuations, we may use the quantum Monte-Carlo simulations.
  In addition, we may investigate the nontrivial effect of the quantum fluctuations, known as the non-stochastic Hamiltonian\cite{Seki2012,Seki2015,Ohzeki2017,Arai2018dy,Okada2019XX}.
  However, longer times are required for each optimization.
  Thus, we consider using the approximate message-passing algorithm depending on the form of the acquisition function\cite{Ohzeki2019JPSJ}.
  However, in the present study, there is no significant difference between thermal and quantum fluctuation to find approximate solution of the acquisition function.
  Thus we may employ other Ising solvers such as the CMOS annealer\cite{Yamaoka2016}, the Fujitsu Digital Annealer\cite{Tsukamoto2017}, TOSHIBA simulated bifurcation algorithm\cite{Goto2019} and the FPGA for performing quantum Monte-Carlo simulation efficiently\cite{Waidyasooriya2019}.
  In addition, the current D-Wave 2000Q performs hybrid computation up to $20,000$ variables on a complete graph.
  By using these solvers and employing BOCS, we compute more complicated optimization problems with a large number of variables beyond our investigations in the present study.
  
  It appears that utilizing a distribution generated from D-Wave devices can improve acquisition functions in the BOCS algorithm, because D-Wave devices return near-optimal solutions within a few microseconds. 
  One of the topics for future work will be finding a surrogate model suited for a distribution generated from D-Wave devices.
  We conclude that difference between thermal and quantum fluctuations is not observed in our results.
  However, the sampling from the D-Wave quantum annealer actually generates the output affected by the quantum fluctuation.
  Therefore, more suitable applications of the quantum device in the framework of Bayesian inference should be considered.
  This is another future research problem.
  
\begin{acknowledgement}
  The authors would like to thank Masamichi J. Miyama and Shuntaro Okada for fruitful discussions.
The present work was financially supported by JSPS KAKENHI
Grant No. 18H03303, 18J20396, 19H01095, 20H02168 and the JST-CREST (No.JPMJCR1402) for Japan Science and Technology Agency, the Next Generation High-Performance Computing Infrastructures and Applications R\&D Program of MEXT and by MEXT-Quantum Leap Flagship Program Grant Number JPMXS0120352009.
\end{acknowledgement}
\bibliographystyle{jpsj}
\bibliography{bib-bocs}

\begin{thebibliography}{10}

\bibitem{oganov2006}
A.~R. Oganov and C.~W. Glass: The Journal of Chemical Physics {\bfseries 124}
  (2006) 244704.

\bibitem{snoek2012}
J.~Snoek, H.~Larochelle, and R.~P. Adams: Proceedings of the 25th International
  Conference on Neural Information Processing Systems - Volume 2, {{NIPS}}'12,
  2012, pp. 2951--2959.

\bibitem{floreano1998}
D.~Floreano and F.~Mondada: Neural Networks {\bfseries 11} (1998) 1461.

\bibitem{jones1998}
D.~R. Jones, M.~Schonlau, and W.~J. Welch: Journal of Global Optimization
  {\bfseries 13} (1998) 455.

\bibitem{baptista2018}
R.~Baptista and M.~Poloczek: In J.~Dy and A.~Krause (eds), {\em Proceedings of
  the 35th International Conference on Machine Learning}, Vol.~80 of {\em
  Proceedings of Machine Learning Research}, July 2018, pp. 462--471.

\bibitem{johnson2011}
M.~W. Johnson, M.~H.~S. Amin, S.~Gildert, T.~Lanting, F.~Hamze, N.~Dickson,
  R.~Harris, A.~J. Berkley, J.~Johansson, P.~Bunyk, E.~M. Chapple, C.~Enderud,
  J.~P. Hilton, K.~Karimi, E.~Ladizinsky, N.~Ladizinsky, T.~Oh, I.~Perminov,
  C.~Rich, M.~C. Thom, E.~Tolkacheva, C.~J.~S. Truncik, S.~Uchaikin, J.~Wang,
  B.~Wilson, and G.~Rose: Nature {\bfseries 473} (2011) 194 EP .

\bibitem{kadowaki1998}
T.~Kadowaki and H.~Nishimori: Phys. Rev. E {\bfseries 58} (1998) 5355.

\bibitem{lucas2014}
A.~Lucas: Frontiers in Physics {\bfseries 2} (2014) 5.

\bibitem{ohzeki2019}
M.~Ohzeki, A.~Miki, M.~J. Miyama, and M.~Terabe: Frontiers in Computer Science
  {\bfseries 1} (2019) 9.

\bibitem{neukart2017}
F.~Neukart, G.~Compostella, C.~Seidel, D.~von Dollen, S.~Yarkoni, and
  B.~Parney: Frontiers in ICT {\bfseries 4} (2017) 29.

\bibitem{amin2018}
M.~H. Amin, E.~Andriyash, J.~Rolfe, B.~Kulchytskyy, and R.~Melko: Physical
  Review X {\bfseries 8} (2018).

\bibitem{Suzuki2005}
S.~Suzuki and M.~Okada: Journal of the Physical Society of Japan {\bfseries 74}
  (2005) 1649.

\bibitem{Morita2008}
S.~Morita and H.~Nishimori: Journal of Mathematical Physics {\bfseries 49}
  (2008).

\bibitem{Ohzeki2011c}
M.~Ohzeki and H.~Nishimori: Journal of Computational and Theoretical
  Nanoscience {\bfseries 8} (2011-06-01T00:00:00) 963.

\bibitem{Santoro2002}
G.~E. Santoro, R.~Marto{\v n}{\'a}k, E.~Tosatti, and R.~Car: Science {\bfseries
  295} (2002) 2427.

\bibitem{Santoro2004}
R.~Marto\ifmmode~\check{n}\else \v{n}\fi{}\'ak, G.~E. Santoro, and E.~Tosatti:
  Phys. Rev. E {\bfseries 70} (2004) 057701.

\bibitem{kirkpatrick1983}
S.~Kirkpatrick, C.~D. Gelatt, and M.~P. Vecchi: Science {\bfseries 220} (1983)
  671.

\bibitem{Ohzeki2018}
M.~Ohzeki, S.~Okada, M.~Terabe, and S.~Taguchi: Scientific Reports {\bfseries
  8} (2018) 9950.

\bibitem{Baldassi2018}
C.~Baldassi and R.~Zecchina: Proceedings of the National Academy of Sciences
  {\bfseries 115} (2018) 1457.

\bibitem{arai2021}
S.~Arai, M.~Ohzeki, and K.~Tanaka: arXiv:2102.08609 [cond-mat,dis-nn]  (2021).

\bibitem{kitai2020}
K.~Kitai, J.~Guo, S.~Ju, S.~Tanaka, K.~Tsuda, J.~Shiomi, and R.~Tamura:
  Physical Review Research {\bfseries 2} (2020).

\bibitem{cai2014}
J.~Cai, W.~G. Macready, and A.~Roy: arXiv:1406.2741 [quant-ph]  (2014).

\bibitem{Okada2019}
S.~Okada, M.~Ohzeki, M.~Terabe, and S.~Taguchi: Scientific Reports {\bfseries
  9} (2019) 2098.

\bibitem{Okada2019Potts}
S.~Okada, M.~Ohzeki, and K.~Tanaka: Journal of the Physical Society of Japan
  {\bfseries 89} (2020) 094801.

\bibitem{Okada2019Potts2}
S.~Okada, M.~Ohzeki, and S.~Taguchi: Scientific Reports {\bfseries 9} (2019)
  13036.

\bibitem{Ohzeki2020}
M.~Ohzeki: Scientific Reports {\bfseries 10} (2020) 3126.

\bibitem{NTTdata2020}
M.~Kuramata, R.~Katsuki, and K.~Nakata: arXiv:2012.10135 [quant-ph]  (2020).

\bibitem{Amin2015}
M.~H. Amin: Phys. Rev. A {\bfseries 92} (2015) 052323.

\bibitem{Bando2020}
Y.~Bando, Y.~Susa, H.~Oshiyama, N.~Shibata, M.~Ohzeki, F.~J. G\'omez-Ruiz,
  D.~A. Lidar, S.~Suzuki, A.~del Campo, and H.~Nishimori: Phys. Rev. Research
  {\bfseries 2} (2020) 033369.

\bibitem{Ohzeki2010a}
M.~Ohzeki: Phys. Rev. Lett. {\bfseries 105} (2010) 050401.

\bibitem{Ohzeki2011}
M.~Ohzeki, H.~Nishimori, and H.~Katsuda: J. Phys. Soc. Jpn. {\bfseries 80}
  (2011) 084002.

\bibitem{Ohzeki2011proc}
M.~Ohzeki and H.~Nishimori: Journal of Physics: Conference Series {\bfseries
  302} (2011) 012047.

\bibitem{Somma2012}
R.~D. Somma, D.~Nagaj, and M.~Kieferov\'a: Phys. Rev. Lett. {\bfseries 109}
  (2012) 050501.

\bibitem{Kadowaki2019}
T.~Kadowaki and M.~Ohzeki: Journal of the Physical Society of Japan {\bfseries
  88} (2019) 061008.

\bibitem{takahashi2018}
C.~Takahashi, M.~Ohzeki, S.~Okada, M.~Terabe, S.~Taguchi, and K.~Tanaka:
  Journal of the Physical Society of Japan {\bfseries 87} (2018) 074001.

\bibitem{Ohzeki2018NOLTA}
M.~Ohzeki, C.~Takahashi, S.~Okada, M.~Terabe, S.~Taguchi, and K.~Tanaka:
  Nonlinear Theory and Its Applications, IEICE {\bfseries 9} (2018) 392.

\bibitem{Carvalho2010}
C.~M. Carvalho, N.~G. Polson, and J.~G. Scott: Biometrika {\bfseries 97} (2010)
  465.

\bibitem{makalic2016}
E.~Makalic and D.~F. Schmidt: IEEE Signal Processing Letters {\bfseries 23}
  (2016) 179.

\bibitem{bhattacharya2016}
A.~Bhattacharya, A.~Chakraborty, and B.~K. Mallick: Biometrika {\bfseries 103}
  (2016) 985.

\bibitem{Hasting1970}
W.~K. Hastings: Biometrika {\bfseries 57} (1970) 97.

\bibitem{Seki2012}
Y.~Seki and H.~Nishimori: Phys. Rev. E {\bfseries 85} (2012) 051112.

\bibitem{Seki2015}
Y.~Seki and H.~Nishimori: Journal of Physics A: Mathematical and Theoretical
  {\bfseries 48} (2015) 335301.

\bibitem{Ohzeki2017}
M.~Ohzeki: Scientific Reports {\bfseries 7} (2017) 41186.

\bibitem{Arai2018dy}
S.~Arai, M.~Ohzeki, and K.~Tanaka: Phys. Rev. E {\bfseries 99} (2019) 032120.

\bibitem{Okada2019XX}
S.~Okada, M.~Ohzeki, and K.~Tanaka: Journal of the Physical Society of Japan
  {\bfseries 88} (2019) 024802.

\bibitem{Ohzeki2019JPSJ}
M.~Ohzeki: Journal of the Physical Society of Japan {\bfseries 88} (2019)
  061005.

\bibitem{Yamaoka2016}
M.~Yamaoka, C.~Yoshimura, M.~Hayashi, T.~Okuyama, H.~Aoki, and H.~Mizuno: IEEE
  Journal of Solid-State Circuits {\bfseries 51} (2016) 303.

\bibitem{Tsukamoto2017}
S.~Tsukamoto, M.~Takatsu, S.~Matsubara, and H.~Tamura: FUJITSU Sci. Tech. J.,
  Vol.~53, 2017, p.~8.

\bibitem{Goto2019}
H.~Goto, K.~Tatsumura, and A.~R. Dixon: Science Advances {\bfseries 5} (2019).

\bibitem{Waidyasooriya2019}
H.~M. Waidyasooriya, M.~Hariyama, M.~J. Miyama, and M.~Ohzeki: The Journal of
  Supercomputing {\bfseries 75} (2019) 5019.

\end{thebibliography}
\end{document}